\documentclass[uplatex,dvipdfmx]{interact}

\usepackage{breqn}
\usepackage{epstopdf}% To incorporate .eps illustrations using PDFLaTeX, etc.
\usepackage{subfigure}% Support for small, `sub' figures and tables
\usepackage{siunitx}
\usepackage{hyperref,bm,amsmath,booktabs,multirow,color,here,graphicx}
\usepackage{caption}
\usepackage{natbib}% Citation support using natbib.sty
\bibpunct[, ]{(}{)}{;}{a}{}{,}% Citation support using natbib.sty
\theoremstyle{plain}% Theorem-like structures provided by amsthm.sty

\theoremstyle{definition}

\theoremstyle{remark}

\begin{document}

%\articletype{Original Paper}
%\articletype{}

\title{Bayesian Fused Lasso Modeling via Horseshoe Prior}

\author{
\name{Yuko Kakikawa\textsuperscript{a}, Kaito Shimamura\textsuperscript{a}, and Shuichi Kawano\textsuperscript{a}\thanks{CONTACT Yuko Kakikawa. Email: kakikawa@ai.lab.uec.ac.jp}}
\affil{\textsuperscript{a}Graduate School of Informatics and Engineering, The University of Electro-Communications, 1-5-1 Chofugaoka, Chofu-shi, Tokyo 182-8585, Japan.}
}

\maketitle

\begin{abstract}
	Bayesian fused lasso is one of the sparse Bayesian methods, which shrinks both regression coefficients and their successive differences simultaneously. In this paper, we propose a Bayesian fused lasso modeling via horseshoe prior. By assuming a horseshoe prior on the difference of successive regression coefficients, the proposed method enables us to prevent over-shrinkage of those differences. We also propose a Bayesian hexagonal operator for regression with shrinkage and equality selection (HORSES) with horseshoe prior, which imposes priors on all combinations of differences of regression coefficients. Simulation studies and an application to real data show that the proposed method gives better performance than existing methods.
\end{abstract}


\section{Introduction}
\label{sec:Introduction}
 Recently, a wide variety of data have come to be used in statistical analysis. Especially, the analysis of high-dimensional data, such as image data and financial data is taking on added significance. To handle these data, it is important to perform variable selection and variable fusion, which correspond to extracting relevant variables and capturing the group structure of data, respectively. To this end, sparse regularization methods such as lasso  \citep{tibshirani1996regression}, fused lasso \citep{tibshirani2005sparsity}, and a hexagonal operator for regression with shrinkage and equality selection (HORSES) \citep{jang2015regression} have been proposed. These methods allow us to execute variable selection and variable fusion by the estimation of regression coefficients. \\
 \indent Meanwhile, many Bayesian approaches to these regularization methods, in which priors on regression coefficients correspond to regularization terms, have also been proposed. For example, \citet{park2008bayesian} proposed Bayesian lasso, which shrinks regression coefficients by assuming they follow a Laplace distribution. Furthermore, \citet{park2008bayesian} developed a Gibbs sampling using a hierarchical expression of the Laplace distribution. \citet{kyung2010penalized} expanded Bayesian lasso by assuming Laplace distributions not only on regression coefficients but also on their successive differences, which is called Bayesian fused lasso. \\
\indent A Laplace prior tends to shrink its targets, such as regression coefficients and their successive differences, too much. To overcome this problem, the Student-$t$ prior and the normal-exponential-gamma (NEG) distribution \citep{griffin2005alternative}, which have heavier tails than a Laplace prior, have also been used. \citet{song2020bayesian} proposed using a Student-$t$ prior to construct Bayesian fusion models. \citet{shimamura2019bayesian} proposed Bayesian fused lasso based on the hierarchical expression of an NEG prior. In addition, a horseshoe prior \citep{carvalho2010horseshoe} is also often used instead of a Laplace prior. A horseshoe prior has an infinite spike at zero and a Cauchy-like tail, which leads to simultaneous weak shrinkage on non-zero elements and strong shrinkage on exactly zero ones. \citet{makalic2015simple} introduced a linear regression model in which a horseshoe prior is assumed on the regression coefficients and developed a simple Gibbs sampler for it. However, these existing methods assume a horseshoe prior on only the regression coefficients.\\
\indent In this paper, we propose Bayesian fused lasso modeling with horseshoe prior under the framework of linear regression models. To formulate the Bayesian model, we assume a Laplace prior on the regression coefficients and a horseshoe prior on their successive differences. We also propose Bayesian HORSES with horseshoe prior, where the horseshoe prior is assumed on every pair of differences of regression coefficients. We develop a Gibbs sampler for the parameters by using the hierarchical expression of the half-Cauchy prior \citep{wand2011mean} shown by \citet{makalic2015simple}. \\
\indent We note that \citet{banerjee2021horseshoe} proposed imposing a horseshoe prior on differences of coefficients. However, \citet{banerjee2021horseshoe} used the model assumed in the one-dimensional fused lasso signal approximation in \citet{friedman2007pathwise}, which is a special case of a linear regression model. In addition, \citet{banerjee2021horseshoe} did not perform variable selection, unlike our proposed method.\\
\indent The remainder of the paper is organized as follows. Section \ref{sec:Bayesian_linear} describes the Bayesian models and introduces sparse Bayesian modelings with horseshoe prior.
In Section \ref{sec:proposed}, we propose Bayesian fused lasso and Bayesian HORSES with horseshoe prior, and then develop Gibbs samplings for them. Section \ref{sec:experiment} presents Monte Carlo simulations and an application to real data to compare our proposed method with existing methods. We conclude our paper in Section \ref{sec:conclusion}.

%%%%%%%%%%%%%%%%%%%%%%%%%%%%%%%%%%%%%%%%%%%%%%%%%%%%%%%%%%%%%%
%%%%%%%%%%%%%%%%%%%%%%%%%%%%%%%%%%%%%%%%%%%%%%%%%%%%%%%%%%%%%%
%%%%%%%%%%%%%%%%%%%%%%%%%%%%%%%%%%%%%%%%%%%%%%%%%%%%%%%%%%%%%%
\section{Sparse Bayesian linear regression modeling}
\label{sec:Bayesian_linear}
In this section, we review Bayesian linear regression, Bayesian lasso, and Bayesian fused lasso. We also describe Bayesian linear regression via horseshoe prior.
\subsection{Preliminaries}
Let $\bm{y}=(y_1,y_2,\ldots,y_n)^{T}$ be an $n$-dimensional vector of the response variable and $\bm{X}=(\bm{x}_{1},\bm{x}_{2},\ldots,\bm{x}_{n})^{T}$ be an $n\times p$ design matrix. A linear regression model is formulated as
\begin{equation}
\bm{y}=\bm{X\beta}+\bm{\epsilon},\label{eq:line}
\end{equation}
where $\bm{\beta}=(\beta_{1},\beta_{2},\ldots,\beta_{p})^{T}$ is a $p$-dimensional regression coefficient vector and $\bm{\epsilon}=(\epsilon_{1},\epsilon_{2},\ldots,\epsilon_{n})^{T}$ is an $n$-dimensional error vector that is distributed as ${\rm N}_{n}(\bm{0}_{n},\sigma^{2}I_{n})$. Here $\bm{0}_{n}$ is an $n$-dimensional vector of zeros, $\sigma^{2}$ is an error variance, and $I_{n}$ is an $n\times n$ identity matrix. Without loss of generality, we suppose that the response variable is centered and the explanatory variable is standardized as follows:
\begin{equation}
\sum_{i=1}^{n}y_{i}=0,\hspace{12pt}\sum_{i=1}^{n}x_{ij}=0,\hspace{12pt}\sum_{i=1}^{n}x_{ij}^{2}=n\hspace{12pt} (j=1,2,\ldots,p).\label{eq:hyojyun}
\end{equation}
\indent Then, the likelihood function is given by
\begin{equation}
f(\bm{y}\mid\bm{X};\bm{\beta},\sigma^{2})=\prod_{i=1}^{n}f(y_{i}\mid\bm{x}_{i};\bm{\beta},\sigma^{2}),\label{eq:yudo}
\end{equation}
where
\begin{equation}
f(y_{i}\mid\bm{x}_{i};\bm{\beta},\sigma^{2})=\frac{1}{\sqrt{2\pi\sigma^{2}}}\exp\left\{-\frac{(y_{i}-\bm{x}_{i}^{T}\bm{\beta})^{2}}{2\sigma^{2}}\right\}.
\label{eq:yudof}
\end{equation}
\subsection{Bayesian lasso}
\citet{tibshirani1996regression} proposed lasso, which performs parameter estimation and variable selection simultaneously in terms of frequentist. He also mentioned that the lasso solution is identical to a posterior mode obtained by imposing the Laplace distribution on the parameter vector $\bm \beta$ as its prior. \\
\indent Based on the perspective of \citet{tibshirani1996regression}, \citet{park2008bayesian} established a Bayesian estimation for lasso. The estimation is called Bayesian lasso. Bayesian lasso considers a conditional Laplace prior in the form
% \citet{park2008bayesian} proposed Bayesian lasso, which shrinks regression coefficients by assuming Laplace distribution on the regression coefficient vector $\bm{\beta}$. Laplace distribution with parameters $\mu$, $b\ (>0)$ is expressed as follows:
% \begin{equation}
%  {\rm Laplace}(x\mid\mu,b)=\frac{b}{2}\exp\left(-b|x-\mu|\right).
%  \label{eq:laplace}
% \end{equation}
% where $\mu$, $b\ (>0)$ are positive parameters.
% In addition, figure \ref{fig:laplace} shows the probability density function of Laplace distribution.
% The prior on regression coefficients $\bm{\beta}$ that \citet{park2008bayesian} supposed is
\begin{equation}
 \begin{split}
   \pi(\bm{\beta}\mid\sigma^{2}){}&={(\sigma^2)}^{-\frac{p}{2}}\prod_{j=1}^{p}{\rm Laplace}\left(\frac{\beta_j}{\sqrt{\sigma^2}}\mid0,\lambda\right)\\
   &=\prod_{j=1}^{p}\frac{\lambda}{2\sqrt{\sigma^{2}}}\exp\left(-\frac{\lambda}{\sqrt{\sigma^{2}}}|\beta_{j}|\right),
   \label{eq:parkprior}
 \end{split}
\end{equation}
where $\lambda \ (>0)$ is a hyper-parameter. Conditioning $\bm{\beta}$ on $\sigma^2$ makes the posterior distribution unimodal (for example, see Appendix A in \citet{park2008bayesian}).
% \begin{figure}[H]
%  \centering
%  \includegraphics[scale=0.6]{laplace2.png}
%  \caption{the probability density function of Laplace distribution}
%  \label{fig:laplace}
% \end{figure}

\indent The prior distribution in \eqref{eq:parkprior} can be rewritten as
\begin{equation}
\frac{\lambda}{2\sqrt{\sigma^{2}}}\exp\left(-\frac{\lambda}{\sqrt{\sigma^{2}}}|\beta|\right)=\int_{0}^{\infty}\frac{1}{\sqrt{2\pi\sigma^{2}\tau^{2}}}\exp\left(-\frac{\beta^{2}}{2\sigma^{2}\tau^{2}}\right)\frac{\lambda^{2}}{2}\exp\left(-\frac{\lambda^{2}}{2}\tau^{2}\right)d\tau^{2}
\label{eq:laplaceseiki}
\end{equation}
by using a scale mixture of normals \citep{andrews1974scale}. This equation means that the Laplace distribution is represented as the convolution of the following two distributions:
\begin{equation}
\begin{split}
&\pi(\bm{\beta}\mid\sigma^{2},\tau_{1}^{2},\ldots,\tau_{p}^{2})=\prod_{j=1}^{p}\frac{1}{\sqrt{2\pi\sigma^{2}\tau_{j}^{2}}}\exp\left(-\frac{\beta_{j}^{2}}{2\sigma^{2}\tau_{j}^{2}}\right),\nonumber
\\&\pi(\tau_{1}^{2},\ldots,\tau_{p}^{2})=\prod_{j=1}^{p}\frac{\lambda^{2}}{2}\exp\left(-\frac{\lambda^{2}}{2}\tau_{j}^{2}\right).
\label{eq:parkpriorAndrew}
\end{split}
\end{equation}
For the parameter $\sigma^2$, the improper prior distribution $\pi(\sigma^2) \propto 1/\sigma^2$ or any inverse gamma distribution for $\sigma^2$ is assumed. Based on the likelihood and the prior distributions, a Gibbs sampling for Bayesian lasso is developed. We omit the Gibbs samplers. For details, we refer the reader to \citet{park2008bayesian}.

\subsection{Bayesian fused lasso}
The fused lasso \citep{tibshirani2005sparsity} encourages sparsity in both the coefficients and their successive differences. \citet{kyung2010penalized} proposed Bayesian fused lasso as a Bayesian counterpart to fused lasso. Bayesian fused lasso assumes a prior distribution for $\bm \beta$ of the following form:
\begin{equation}
\pi(\bm{\beta}\mid\sigma^{2})\propto(\sigma^{2})^{-\frac{2p-1}{2}}\exp\left(-\frac{\lambda_{1}}{\sigma}\sum_{j=1}^{p}|\beta_{j}|-\frac{\lambda_{2}}{\sigma}\sum_{j=2}^{p}|\beta_{j}-\beta_{j-1}|\right),
\label{eq:bayesianfusedprior}
\end{equation}
where $\lambda_1$ and $\lambda_2$ are positive hyper-parameters. Similar to Bayesian lasso, a scale mixture of normals is applied. Then the prior distribution \eqref{eq:bayesianfusedprior} is transformed into
\begin{equation}
\begin{split}
\pi(\bm{\beta}\mid\sigma^{2})
\propto{}&(\sigma^{2})^{-\frac{2p-1}{2}}\prod_{j=1}^{p}\int \frac{1}{\sqrt{2\pi\tau_{j}^{2}}}\exp\left(-\frac{\beta_{j}^{2}}{2\sigma^{2}\tau_{j}^{2}}\right)\frac{\lambda_{1}^{2}}{2}\exp\left(-\frac{\lambda_{1}^{2}}{2}\tau_{j}^{2}\right)d\tau_{j}^{2}
\\&\times\prod_{j=2}^{p}\int\frac{1}{\sqrt{2\pi\tilde{\tau}_{j}^{2}}}\exp\left\{-\frac{(\beta_{j}-\beta_{j-1})^2}{2\sigma^{2}\tilde{\tau}_{j}^{2}}\right\}\frac{\lambda_{2}^{2}}{2}\exp\left(-\frac{\lambda_{2}^{2}}{2}\tilde{\tau}_{j}^{2}\right)d\tilde{\tau}_{j}^{2}.
\end{split}
\label{eq:bayesianfusedpriorkaki}
\end{equation}
Using this hierarchical relationship, \citet{kyung2010penalized} developed a Gibbs sampling for Bayesian fused lasso. \\
\indent To perform the fully Bayesian estimation, \citet{kyung2010penalized} further assumed the gamma distribution
for the hyper-parameters $\lambda_{1}$ and $\lambda_{2}$ as
\begin{equation}
\begin{split}
{}&\lambda_{1}^2\sim {\rm Ga}(r_{1},\delta_{1}),\\
&	\lambda_{2}^2\sim {\rm Ga}(r_{2},\delta_{2}),
\end{split}
\end{equation}
where $r_{1}, r_{2},\delta_{1}$, and $\delta_{2}$ are positive hyper-parameters. Here, the probability density function of the gamma distribution is given by
\begin{equation}
 {\rm Ga}(x\mid m,c)=\frac{c^{m}}{\Gamma(m)}x^{m-1}\exp(-cx)\hspace{50pt}(x\geq0),\nonumber
\end{equation}
where $m$ is a shape parameter and $c$ is a rate parameter, both taking positive values. In \citet{kyung2010penalized}, $r_{1}=1, r_{2}=1,\delta_{1}=10$, and $\delta_{2}=10$ are used because the gamma distribution is relatively flat with these parameter values. We omit the full conditional posteriors and the Gibbs samplers. For details, we refer the reader to \citet{kyung2010penalized}.\\
\indent Next, we explain HORSES by \citet{jang2015regression}. HORSES was proposed as an extension of fused lasso; HORSES imposes an $L_1$ penalty on all combinations of differences of regression coefficients. In the Bayesian framework, this corresponds to assuming a Laplace prior of the form
\begin{equation}
	\begin{split}
		\pi(\bm{\beta}\mid\sigma^{2})
		 \propto{}&
		 (\sigma^{2})^{-\frac{p^{2}+p}{4}}\prod_{j=1}^{p}{\rm Laplace}\left(\frac{\beta_{j}}{\sqrt{\sigma^{2}}}\mid0,\lambda_{1}\right)
		\\&\times\prod_{j>k}^{p}{\rm Laplace}\left(\frac{\beta_{j}-\beta_{k}}{\sqrt{\sigma^{2}}}\mid0,\lambda_{2}\right)
  \end{split}
\end{equation}
for regression coefficients $\bm \beta$. Note that HORSES is also known as generalized fused lasso \citep{she2010sparse}.\\
\indent The Laplace distribution shrinks all of the regression coefficients to the same extent. \citet{shimamura2019bayesian} proposed Bayesian fused lasso and Bayesian HORSES with NEG prior. This method assumes a Laplace prior on the regression coefficients and an NEG prior on their differences. Because an NEG prior has two properties, a spike at zero and extreme flatness of its tail, the method with an NEG prior has the advantage that exactly identical regression coefficients tend to be estimated as identical, while different regression coefficients tend to be estimated as different.
%as
% \begin{equation}
% 	\begin{split}
% 	\pi(\bm{\beta}|\sigma^{2})
% 	 ={}&
% 	 (\sigma^{2})^{-\frac{2p-1}{2}}\prod_{j=1}^{p}\rm Laplace\left(\frac{\beta_{j}}{\sqrt{\sigma^{2}}}|\lambda_{1}\right)
% 	\\&\times\prod_{j=2}^{p}{\rm NEG}\left(\frac{\beta_{j}-\beta_{j-1}}{\sqrt{\sigma^{2}}}|\lambda_{2},\gamma_{2}\right)
% 	\end{split}
% \end{equation}
% where $\lambda_{1}\ (>0), \lambda_{2}\ (>0),\gamma_{2}\ (>0)$ are hyper-parameters.

% \begin{equation}
%  {\rm Ga}(x\mid m,c)=\frac{c^{m}}{\Gamma(m)}x^{m-1}\exp(-cx)\hspace{50pt}(x\geq0)
% \end{equation}
% where $m,c$ are positive parameters.\\
\subsection{Bayesian linear regression model with horseshoe prior}
\citet{makalic2015simple} proposed the following Bayesian linear regression model:
\begin{equation}
\begin{split}
\bm{y}\mid\bm{X},\bm{\beta},\sigma^{2}&\sim{\rm N}_{n}(\bm{X\beta},\sigma^{2}\bm{I}_{n}),
\\\beta_{j}\mid\lambda_{j}^{2},\tau^{2},\sigma^{2}&\sim{\rm N}(0,\lambda_{j}^{2}\tau^{2}\sigma^{2}),
\\
\sigma^{2}&\sim\sigma^{-2}d\sigma^{2},
\\
\lambda_{j}&\sim{\rm C^{+}}(0,1),
\\
\tau&\sim{\rm C^{+}}(0,1).
\end{split}
\label{eq:makalicprior}
\end{equation}
Here, ${\rm C^{+}}(0,a)\ (a>0)$ is a half-Cauchy distribution, which has the following density function:
\begin{equation}
p(x)=\frac{2a}{\pi \left(x^{2}+a^{2}\right)}, \hspace{50pt}(x > 0).\ \ \ \ \ \nonumber
\label{eq:halfcauchy}
\end{equation}
The hierarchies of priors in \eqref{eq:makalicprior} represent the horseshoe prior proposed in \citet{carvalho2010horseshoe}. In the model with horseshoe prior, the half-Cauchy prior distribution is assumed on hyper-parameters $\lambda_{j}$ and $\tau$. Hyper-parameter $\lambda_{j}$ adjusts the level of local shrinkage for regression coefficient $\beta_{j}$, while hyper-parameter $\tau$ determines the degree of global shrinkage for all regression coefficients. Owing to having these two types of hyper-parameters, the horseshoe prior simultaneously enjoys a heavy tail and infinitely tall spike at zero. These properties induce exactly identical regression coefficients to tend to be estimated as identical, while different regression coefficients tend to be estimated as different.\\
\indent To develop a Gibbs sampling for the parameters, \citet{makalic2015simple} used a hierarchical expression of the half-Cauchy distribution \citep{wand2011mean}, which means that $x$ follows ${\rm C}^{+}(0,A)$ when $x^{2}$ and $a$ have the following priors:
\begin{equation}
x^{2}\mid a\sim{\rm IG}\left(\frac{1}{2},\frac{1}{a}\right),\ \ \ a\sim{\rm IG}\left(\frac{1}{2},\frac{1}{A^{2}}\right),
\label{eq:halfcauchyxa}
\end{equation}
where $A$ is a positive constant. Using \eqref{eq:halfcauchyxa}, the priors of the model \eqref{eq:makalicprior} can be expressed as follows:
\begin{equation}
\begin{split}
\bm{y}\mid\bm{X},\bm{\beta},\sigma^{2}&\sim{\rm N}_{n}(\bm{X\beta},\sigma^{2}\bm{I}_{n}),
\\\beta_{j}\mid\lambda_{j}^{2},\tau^{2},\sigma^{2}&\sim{\rm N}(0,\lambda_{j}^{2}\tau^{2}\sigma^{2}),
\\
\sigma^{2}&\sim\sigma^{-2}d\sigma^{2},
\\
\lambda_{j}^{2}\mid\nu_{j}&\sim{\rm IG}\left(\frac{1}{2},\frac{1}{\nu_{j}}\right),
\\
\tau^{2}\mid\xi&\sim{\rm IG}\left(\frac{1}{2},\frac{1}{\xi}\right),
\\
\nu_{1},\ldots,\nu_{p},\xi&\sim\rm{IG}\left(\frac{1}{2},1\right).
\end{split}
\label{eq:makalicpriorkaki}
\end{equation}
We omit the full conditional posteriors and the Gibbs samplers. For details, we refer the reader to \citet{makalic2015simple} and \citet{nalenz2018tree}.
\section{Proposed method}
\label{sec:proposed}
In this section, we propose the Bayesian linear regression modeling, which assumes the horseshoe prior on successive differences of regression coefficients. We also extend this approach to HORSES.
\subsection{Bayesian fused lasso with horseshoe prior}
We propose assuming a Laplace prior on regression coefficients and a horseshoe prior on their successive differences as follows:
\begin{equation}
\begin{split}
\pi(\bm{\beta}\mid\sigma^{2})
 \propto{}&(\sigma^{2})^{-\frac{p}{2}}\prod_{j=1}^{p}{\rm Laplace}\left(\frac{\beta_{j}}{\sqrt{\sigma^{2}}}\mid\tilde{\lambda}_{1}\right)
\\&\times\prod_{j=2}^{p} \iint\frac{1}{\sqrt{2\pi\lambda_{j}^{2}\tilde{\tau}^{2}\sigma^{2}}}\exp\left\{-\frac{(\beta_{j}-\beta_{j-1})^2}{2\lambda_{j}^{2}\tilde{\tau}^{2}\sigma^{2}}\right\}
\frac{2}{\pi(1+\lambda_{j}^{2})}\frac{2}{\pi(1+\tilde{\tau}^{2})}d\lambda_{j}^{2}d\tilde
{\tau}^{2},
\end{split}
\label{eq:yukoprior}
\end{equation}
By assuming the prior \eqref{eq:yukoprior}, small differences between successive regression coefficients are largely shrunk, while large differences are not much shrunk.\\
\indent Using a scale mixture of normals \citep{andrews1974scale}, the prior \eqref{eq:yukoprior} can be expressed as follows:
\begin{equation}
\begin{split}
\pi(\bm{\beta}\mid\sigma^{2})\propto{}&%4行目
\int \ldots \int (\sigma^{2})^{-\frac{2p-1}{2}}(\tilde{\tau}^{2})^{-\frac{p-1}{2}}\pi(\tilde{\tau}^{2}\mid\xi)\pi(\xi)\prod_{j=1}^{p}(\tau_{j}^{2})^{-\frac{1}{2}}\prod_{j=2}^{p}(\lambda_{j}^{2})^{-\frac{1}{2}}\exp\left(-\frac{1}{2\sigma^{2}}\bm{\beta}^{T}\bm{B}^{-1}\bm{\beta}\right)
 \\& \times\prod_{j=1}^{p}\pi(\tau_{j}^{2})\prod_{j=2}^{p}\pi(\lambda_{j}^{2}\mid\nu_{j})\prod_{j=2}^{p}\pi(\nu_{j})d\tilde{\tau}^{2}d\xi\prod_{j=1}^{p}d\tau_{j}^{2}\prod_{j=2}^{p}d\lambda_{j}^{2}\prod_{j=2}^{p}d\nu_{j}.
\end{split}
\label{eq:yukoprior_rewrite}
\end{equation}
% In addition, we assume the prior on $\sigma^2$ as follows:
% \begin{equation}
%  \sigma^2\sim {\rm IG}(\frac{\nu_0}{2},\frac{\eta_0}{2}).
%  \label{eq:sigma}
% \end{equation}
Here, the inverse of matrix $\bm{B}$ is represented by
\begin{equation}
\bm{B}^{-1}=\left(\begin{array}{cccccc}
\frac{1}{\tau_{1}^{2}}+\frac{1}{\lambda_{2}^{2}\tilde{\tau}^{2}}&-\frac{1}{\lambda_{2}^{2}\tilde{\tau}^{2}}&0&\ldots&0&0\\
-\frac{1}{\lambda_{2}^{2}\tilde{\tau}^{2}}&\frac{1}{\tau_{2}^{2}}+\frac{1}{\lambda_{2}^{2}\tilde{\tau}^{2}}+\frac{1}{\lambda_{3}^{2}\tilde{\tau}^{2}}&-\frac{1}{\lambda_{3}^{2}\tilde{\tau}^{2}}&\ldots&0&0\\
0&-\frac{1}{\lambda_{3}^{2}\tilde{\tau}^{2}}&\frac{1}{\tau_{3}^{2}}+\frac{1}{\lambda_{3}^{2}\tilde{\tau}^{2}}+\frac{1}{\lambda_{4}^{2}\tilde{\tau}^{2}}&\ldots&0&0\\
\vdots&\vdots&\vdots&\ddots&\vdots&\vdots\\
0&0&0&\ldots&\frac{1}{\tau_{p-1}^{2}}+\frac{1}{\lambda_{p-1}^{2}\tilde{\tau}^{2}}+\frac{1}{\lambda_{p}^{2}\tilde{\tau}^{2}}&-\frac{1}{\lambda_{p}^{2}\tilde{\tau}^{2}}\\
0&0&0&\ldots&-\frac{1}{\lambda_{p}^{2}\tilde{\tau}^{2}}&\frac{1}{\tau_{p}^{2}}+\frac{1}{\lambda_{p}^{2}\tilde{\tau}^{2}}
\end{array}\right).
\end{equation}
The detailed calculation of \eqref{eq:yukoprior_rewrite} is given in Appendix A. Therefore, the priors on $\bm{\beta},\tau_{1}^{2},\ldots,\tau_{p}^{2},\tilde{\tau}^{2},\lambda_{2}^{2},\ldots,\lambda_{p}^{2},\nu_{2},\ldots,\nu_{p},\xi$ are given by
\begin{equation}
\begin{split}
\bm{\beta}\mid\sigma^{2},\tau_{1}^{2},\ldots,\tau_{p}^{2},\tilde{\tau}^{2},\lambda_{2}^{2},\ldots,\lambda_{p}^{2}&\sim {\rm N}_{p}({\bm 0}_{p},\sigma^{2}\bm{B}),
\\
\tau_{j}^{2}&\sim{\rm EXP}\left(\frac{\tilde{\lambda}_{1}^{2}}{2}\right),
\\
\tilde{\tau}^{2}\mid\xi&\sim{\rm IG}\left(\frac{1}{2},\frac{1}{\xi}\right),
\\
\lambda_{j}^{2}\mid\nu_{j}&\sim{\rm IG}\left(\frac{1}{2},\frac{1}{\nu_{j}}\right),\hspace{24pt}(j=2,\ldots,p),
\\
\xi,\nu_{j}&\sim{\rm IG}\left(\frac{1}{2},1\right),\hspace{24pt}(j=2,\ldots,p),\nonumber
\end{split}
\end{equation}
where ${\rm EXP}(x\mid d)$ is an exponential prior with density function
\begin{equation}
 {\rm EXP}(x\mid d)=d\exp(-dx),\hspace{50pt}(x\geq0).
\end{equation}
Here $d$ is positive. In addition, we assume the priors on $\sigma^{2}$ and $\tilde{\lambda}_{1}^{2}$ as
\begin{equation}
	\begin{split}
		\sigma^{2}{}&\sim {\rm IG}(\frac{\nu_{0}}{2},\frac{\eta_{0}}{2}),\\
		\tilde{\lambda}_{1}^{2}&\sim {\rm Ga}(r_{1},\delta_{1}).
	\end{split}
	\label{eq:sigmaandlambda}
\end{equation}
\indent By using the likelihood and the priors for the parameters, we can obtain the full conditional distributions as follows:
\begin{flalign}
\bm{\beta}\mid\bm{y},\bm{X},\sigma^{2},\tau_{1}^{2},\ldots,\tau_{p}^{2},\tilde{\tau}^{2},\lambda_{2}^{2},\ldots,\lambda_{p}^{2}&\sim{\rm N}_{p}(\bm{A}^{-1}\bm{X}^{T}\bm{y},\sigma^{2}\bm{A}^{-1})\nonumber
\\
\bm{A}&=\bm{X}^{T}\bm{X}+\bm{B}^{-1},&\nonumber
\\
\sigma^{2}\mid\bm{y},\bm{X},\bm{\beta},\tau_{1}^{2},\ldots,\tau_{p}^{2},\tilde{\tau}^{2},\lambda_{2}^{2},\ldots,\lambda_{p}^{2}&\sim{\rm IG}\left(\frac{n_{1}}{2},\frac{s_{1}}{2}\right)\nonumber
\\
n_{1}&=n+2p-1+\nu_{0},&\nonumber
\\
s_{1}&=(\bm{y}-\bm{X\beta})^{T}(\bm{y}-\bm{X\beta})+\bm{\beta}^{T}\bm{B}^{-1}\bm{\beta}+\eta_{0},&\nonumber
\\
\frac{1}{\tau_{j}^{2}}\mid\beta_{j},\sigma^{2},\tilde{\lambda}_{1}^{2}&\sim{\rm IGauss}\left(\sqrt{\frac{\sigma^{2}\tilde{\lambda}_{1}^{2}}{\beta_{j}^{2}}},\tilde{\lambda}_{1}^{2}\right),\nonumber
\\
\tilde{\lambda}_{1}^{2}\mid\tau^{2}_{1},\ldots,\tau^{2}_{p}&\sim {\rm Ga}\left(p+r_{1},\frac{1}{2}\sum_{j=1}^{p}\tau_{j}^{2}+\delta_{1}\right),\nonumber
\\
\tilde{\tau}^{2}\mid\beta_{1},\ldots,\beta_{p},\sigma^{2},\lambda_{2}^{2},\ldots,\lambda_{p}^{2},\xi&\sim{\rm IG}\left(\frac{p}{2},\frac{1}{2\sigma^{2}}\sum^{p}_{j=2}\frac{(\beta_{j}-\beta_{j-1})^{2}}{\lambda_{j}^{2}}+\frac{1}{\xi}\right),\nonumber
\\
\lambda_{j}^{2}\mid\beta_{j},\beta_{j-1},\sigma^{2},\tilde{\tau}^{2},\nu_{j}&\sim{\rm IG}\left(1,\frac{(\beta_{j}-\beta_{j-1})^{2}}{2\sigma^{2}\tilde{\tau}^{2}}+\frac{1}{\nu_{j}}\right),\nonumber
\\
\nu_{j}\mid\lambda_{j}^{2}&\sim{\rm IG}\left(1,\frac{1}{\lambda_{j}^{2}}+1\right),\nonumber
\\
\xi\mid\tilde{\tau}^{2}&\sim{\rm IG}\left(1,\frac{1}{\tilde{\tau}^{2}}+1\right).\nonumber
\end{flalign}
By using the full conditional distributions, we can perform the Gibbs sampling.
\subsection{Bayesian HORSES with horseshoe prior}
Next, we propose assuming a Laplace prior on the regression coefficients and a horseshoe prior on all combinations of their differences as follows:
\begin{equation}
\begin{split}
\pi(\bm{\beta}\mid\sigma^{2})
 \propto{}&(\sigma^{2})^{-\frac{p}{2}}\prod_{j=1}^{p}{\rm Laplace}\left(\frac{\beta_{j}}{\sqrt{\sigma^{2}}}\mid\tilde{\lambda}_{1}\right)
\\&\times\prod_{j>k} \int\frac{1}{\sqrt{2\pi\lambda_{j,k}^{2}\tilde{\tau}^{2}\sigma^{2}}}\exp\left\{-\frac{(\beta_{j}-\beta_{k})^2}{2\lambda_{j,k}^{2}\tilde{\tau}^{2}\sigma^{2}}\right\}
\frac{2}{\pi(1+\lambda_{j,k}^{2})}\frac{2}{\pi(1+\tilde{\tau}^{2})}d\lambda_{j,k}^{2}d\tilde{\tau}^{2}.
\end{split}
\label{eq:yukoprior_generalized}
\end{equation}
Therefore, the priors on $\bm{\beta},\tau_{1}^{2},\ldots,\tau_{p}^{2},\tilde{\tau}^{2},\lambda_{1,2}^{2},\ldots,\lambda_{p-1,p}^{2},\nu_{1,2}\ldots,\nu_{p-1,p}$, and $\xi$ are given by
\begin{equation}
\begin{split}
\bm{\beta}\mid\sigma^{2},\tau_{1}^{2}\ldots,\tau_{p}^{2},\tilde{\tau}^{2},\lambda_{1,2}^{2},\ldots,\lambda_{p-1,p}^{2}&\sim {\rm N}_{p}({\bm 0}_{p},\sigma^{2}\bm{B}),
\\
\tau_{j}^{2}&\sim{\rm EXP}\left(\frac{\tilde{\lambda}_{1}^{2}}{2}\right),
\\
\lambda_{j,k}^{2}\mid\nu_{j,k}&\sim{\rm IG}\left(\frac{1}{2},\frac{1}{\nu_{j,k}}\right),
\\
\nu_{j,k}&\sim{\rm IG}\left(\frac{1}{2},1\right),\nonumber
\end{split}
\end{equation}
where $\lambda_{j,k}^{2}=\lambda_{k,j}^{2}$, $\nu_{j,k}^{2}=\nu_{k,j}^{2}$ and the $(i,j)$-element of $\bm{B}^{-1}$ is represented as
\begin{equation}
{\bm{B}^{-1}}_{(i,j)}=\left\{
\begin{split}
	{}&\frac{1}{\tau_{j}^{2}}+\frac{1}{\tilde{\tau}^{2}}\sum_{l\neq i}\frac{1}{\lambda_{i,l}^{2}},\hspace{12pt}(i=j),\\
	&-\frac{1}{\lambda_{i,j}^{2}},\hspace{12pt}(i\neq j).
\end{split}
\right.
\end{equation}
\indent By assuming an inverse gamma prior on $\sigma^2$ in \eqref{eq:sigmaandlambda}, the full conditional distributions are represented as
\begin{flalign}
\bm{\beta}\mid\bm{y},\bm{X},\sigma^{2},\tau_{1}^{2},\ldots,\tau_{p}^{2},\tilde{\tau}^{2},\lambda_{1,2}^{2},\ldots,\lambda_{p-1,p}^{2}&\sim{\rm N}_{p}(\bm{A}^{-1}\bm{X}^{T}\bm{y},\sigma^{2}\bm{A}^{-1}),\nonumber
\\
\bm{A}&=\bm{X}^{T}\bm{X}+\bm{B}^{-1},&\nonumber
\\
\sigma^{2}\mid\bm{y},\bm{X},\bm{\beta},\tau_{1}^{2},\ldots,\tau_{p}^{2},\tilde{\tau}^{2},\lambda_{1,2}^{2},\ldots,\lambda_{p-1,p}^{2}&\sim{\rm IG}\left(\frac{n_{1}}{2},\frac{s_{1}}{2}\right),\nonumber
\\
n_{1}&=n+p(p+1)/2+\nu_{0},&\nonumber
\\
s_{1}&=(\bm{y}-\bm{X\beta})^{T}(\bm{y}-\bm{X\beta})+\bm{\beta}^{T}\bm{B}^{-1}\bm{\beta}+\eta_{0},&\nonumber
\\
\frac{1}{\tau_{j}^{2}}\mid\beta_{j},\sigma^{2},\tilde{\lambda}_{1}^{2}&\sim{\rm IGauss}\left(\sqrt{\frac{\sigma^{2}\tilde{\lambda}_{1}^{2}}{\beta_{j}^{2}}},\tilde{\lambda}_{1}^{2}\right),\nonumber
\\
\tilde{\lambda}_{1}^{2}\mid\tau^{2}_{1},\ldots,\tau^{2}_{p}&\sim {\rm Ga}\left(p+r_{1},\frac{1}{2}\sum_{j=1}^{p}\tau_{j}^{2}+\delta_{1}\right),\nonumber
\\
\lambda_{j,k}^{2}\mid\beta_{j},\beta_{k},\sigma^{2},\tilde{\tau}^{2},\nu_{j,k}&\sim{\rm IG}\left(1,\frac{(\beta_{j}-\beta_{k})^{2}}{2\sigma^{2}\tilde{\tau}^{2}}+\frac{1}{\nu_{j,k}}\right),\nonumber
\\
\nu_{j,k}\mid\lambda_{j,k}^{2}&\sim{\rm IG}\left(1,\frac{1}{\lambda_{j,k}^{2}}+1\right).\nonumber%\hspace{24pt}(j=1,\ldots,p,\hspace{12pt}j>k)
\end{flalign}
By using the full conditional distributions, we can perform the Gibbs sampling for Bayesian HORSES.\\
\indent Note that the hyper-parameter $\tilde{\tau}^2$ is treated as a tuning parameter. The value of the tuning parameter is selected by any model selection criterion such as the widely applicable information criterion (WAIC) \citep{watanabe2010equations}.
\newpage
\section{Numerical studies}
\label{sec:experiment}
In this section, we compare the proposed method with existing methods through Monte Carlo simulations and show its effectiveness. In addition, we apply the proposed method to soil data \citep{bondell2008simultaneous}.
\subsection{Monte Carlo simulation}
We conducted Monte Carlo simulations with artificial data generated from the true model
\begin{equation}
\bm{y}=\bm{X\beta^{*}}+\bm{\epsilon},
\end{equation}
where $\bm{\beta}^{*}$ is the $p$-dimensional true coefficient vector and the error vector $\bm{\epsilon}$ is distributed normally as ${\rm N}_{n}(\bm{0}_{n},\sigma^2\bm{I}_n)$. In addition, $\bm{x}_{i}$ $(i=1,2,\ldots,n)$ is distributed according to the multivariate normal distribution ${\rm N}_{p}(\bm{0}_p, \Sigma)$.\\
\indent We considered the following settings:\\
\newline
Case 1: $\bm{\beta}^{*}=\bm{\beta}_{1}^{*}$ or $\bm{\beta}_{2}^{*}$, $\Sigma_{ii}=1$, $\Sigma_{ij}=0.5,\hspace{12pt}(i\neq j)$,\\
Case 2: $\bm{\beta}^{*}=\bm{\beta}_{1}^{*}$ or $\bm{\beta}_{2}^{*}$, $\Sigma_{ii}=1$, $\Sigma_{ij}=0.5^{|i-j|},\hspace{12pt}(i\neq j)$,\\
Case 3: $\bm{\beta}^{*}=(\bm{3.0}_{5}^{T}, \bm{-1.5}_{5}^{T}, \bm{1.0}_{5}^{T}, \bm{2.0}_{5}^{T},\bm{0.0}_{p-20}^{T})^{T}$, $\Sigma_{ii}=1$, $\Sigma_{ij}=0.5,\ \hspace{12pt}(i\neq j)$,\\
Case 4: $\bm{\beta}^{*}=(\bm{3.0}_{5}^{T}, \bm{-1.5}_{5}^{T}, \bm{1.0}_{5}^{T}, \bm{2.0}_{5}^{T},\bm{0.0}_{p-20}^{T})^{T}$, $\Sigma_{ii}=1$, $\Sigma_{ij}=0.5^{|i-j|},\hspace{12pt}(i\neq j)$,\\
\newline
where $\Sigma_{ij}$ is the $(i,j)$-th element of $\Sigma$. For each case, we considered $\sigma=0.5,1.5$. We simulated Cases 1 and 2 with $\bm{\beta}_{1}^{*}=(\bm{0.0}_{5}^{T}, \bm{1.0}_{5}^{T}, \bm{0.0}_{5}^{T},  \bm{1.0}_{5}^{T})^{T}$ and $\bm{\beta}_{2}^{*}=(\bm{0.0}_{5}^{T}, \bm{2.0}_{5}^{T}, \bm{0.0}_{5}^{T},  \bm{2.0}_{5}^{T})^{T}$. We considered $n=50$ for Cases 1 and 2 and $n=30,50$ for Cases 3 and 4. Therefore, Cases 1 and 2 correspond to $n>p$ cases, whereas Cases 3 and 4 correspond to $n=p$ and $n<p$ cases. We simulated 100 datasets for each case. Cases 1 and 2 are according to example 1 in \citet{shen2010grouping}, whereas Cases 3 and 4 are respectively according to examples 2 and 3 in the same reference. For each case, the Gibbs sampling was run with 5,000 iterations (where we discarded the first 2,000 iterations as burn-in).\\
\indent We compared Bayesian fused lasso with horseshoe prior (BFH) to Bayesian
fused lasso (BFL) and Bayesian fused lasso with NEG prior (BFNEG). For BFNEG, the shape parameter in the Gamma distribution in the NEG prior was set to 0.5 according to the simulation study in \citet{griffin2011bayesian}, while the rate parameter was selected by WAIC.\\
\indent To evaluate the accuracy of the estimation of regression coefficients, we used mean squared error (MSE):
\begin{equation}
 {\rm MSE}=\frac{1}{100}\sum_{k=1}^{100}(\hat{\bm{\beta}}^{(k)}-\bm{\beta}^{*})^{T}(\hat{\bm{\beta}}^{(k)}-\bm{\beta}^{*}),
\end{equation}
where $\hat{\bm{\beta}}^{(k)}=(\hat{\beta}_{1}^{(k)},\ldots,\hat{\beta}_{p}^{(k)})^{T}$ is the regression coefficient vector estimated from the $k$-th dataset. We also computed ${\rm MSE_{diff}}$, expressed as
\begin{equation}
{\rm MSE_{diff}}=\frac{1}{100}\sum_{k=1}^{100}(\hat{\bm{\beta}}_{\rm diff}^{(k)}-\bm{\beta}_{\rm diff}^{*})^{T}(\hat{\bm{\beta}}_{\rm diff}^{(k)}-\bm{\beta}_{\rm diff}^{*}),
\end{equation}
where $\bm{\beta}_{\rm diff}^{*}$ is a vector of the non-zero differences of the true regression coefficients and $\hat{\bm{\beta}}_{\rm diff}^{(k)}$ is the estimated value of $\bm{\beta}_{\rm diff}^{*}$ from the $k$-th dataset. ${\rm MSE_{diff}}$ is an index to assess how close the differences of estimated regression coefficients which are not zero are to the true differences. For example, regression coefficients for Case 1 are given by $\bm{\beta}^{*}=(\bm{0.0}_{5}^{T},\bm{2.0}_{5}^{T},\bm{0.0}_{5}^{T},\bm{2.0}_{5}^{T})^{T}$ and the non-zero successive differences are between the 5th and 6th, 10th and 11th, and 15th and 16th elements of $\bm{\beta}^{*}$. Then, ${\rm MSE_{diff}}$ is calculated as follows:
\begin{dmath}
		{\rm MSE_{diff}}
		=\frac{1}{100}\sum_{k=1}^{100}\left[\left\{(\hat{\beta}_{6}^{(k)}-\hat{\beta}_{5}^{(k)})-(\beta_{6}^{*}-\beta_{5}^{*})\right\}^{2}+\left\{(\hat{\beta}_{11}^{(k)}-\hat{\beta}_{10}^{(k)})-(\beta_{11}^{*}-\beta_{10}^{*})\right\}^{2}+\left\{(\hat{\beta}_{16}^{(k)}-\hat{\beta}_{15}^{(k)})-(\beta_{16}^{*}-\beta_{15}^{*})\right\}^{2}\right].
\end{dmath}
 In addition, we computed prediction squared error
\begin{equation}
 {\rm PSE}=\frac{1}{100}\sum_{k=1}^{100}(\hat{\bm{\beta}}^{(k)}-\bm{\beta}^{*})^{T}\Sigma(\hat{\bm{\beta}}^{(k)}-\bm{\beta}^{*})
\end{equation}
to evaluate the accuracy of prediction.
\begin{table}[H]
 \centering
\caption{MSE (standard deviation), ${\rm MSE_{diff}}$, and PSE for Case 1. Bold font indicates smallest value among BFL, BFNEG, and BFH.}
\label{tab:case1}
 \begin{tabular}{c|cccccccc}
 \hline
 \multicolumn{2}{c}{}&\multicolumn{3}{c}{$\sigma=0.5$} & &\multicolumn{3}{c}{$\sigma=1.5$}\\
 \cline{3-5}\cline{7-9}
\multicolumn{2}{c}{}&MSE&${\rm MSE_{diff}}$&PSE& &MSE&${\rm MSE_{diff}}$&PSE\\
\multicolumn{2}{c}{}&(sd)&(sd)&(sd)& &(sd)&(sd)&(sd)\\
\hline
%\\
\multirow{6}{*}{$\bm{\beta}_{1}^{*}$}&\multirow{2}{*}{BFL} &0.329&0.118&0.325&&2.056&0.723&1.226\\
&&(0.146)&(0.102)&(0.277)&&(0.902)&(0.561)&(0.589)\\
\cline{3-9}
%&& (2.202)&&&(0.360)&\\
%&BFNEG&&&\\
&\multirow{2}{*}{BFNEG}&0.219&\textbf{0.089}&\textbf{0.275}&&1.444&\textbf{0.614}&0.929\\
&&(0.082)&(0.066)&(0.215)&&(0.632)&(0.414)&(0.480)\\
\cline{3-9}
&\multirow{2}{*}{BFH}&\textbf{0.187}&0.098&0.283&&\textbf{1.094}&0.760&\textbf{0.770}\\
&&(0.105)&(0.096)&(0.278)&&(0.571)&(0.501)&(0.452)\\
%&& (1.055)&&&(0.127)&\\
\hline
\multirow{6}{*}{$\bm{\beta}_{2}^{*}$}&\multirow{2}{*}{BFL} &0.623&0.205&0.945&&2.459&0.911&1.893\\
&&(0.307)&(0.173)&(1.076)&&(1.092)&(0.786)&(1.243)\\
\cline{3-9}
%&& (2.202)&&&(0.360)&\\
%&BFNEG&&&\\
&\multirow{2}{*}{BFNEG}&0.504&\textbf{0.173}&\textbf{0.909}&&1.720&\textbf{0.773}&1.438\\
&&(0.252)&(0.144)&(0.993)&&(0.806)&(0.702)&(0.975)\\
\cline{3-9}
&\multirow{2}{*}{BFH}&\textbf{0.454}&0.200&0.982&&\textbf{1.306}&0.791&\textbf{1.429}\\
&&(0.276)&(0.196)&(1.079)&&(0.746)&(0.752)&(1.199)\\
%&& (1.055)&&&(0.127)&\\
\hline
 \end{tabular}
\end{table}

\begin{table}[H]
 \centering
\caption{MSE (standard deviation), ${\rm MSE_{diff}}$, and PSE for Case 2. Bold font indicates smallest value among BFL, BFNEG, and BFH.}
\label{tab:case2}
 \begin{tabular}{c|cccccccc}
 \hline
 \multicolumn{2}{c}{}&\multicolumn{3}{c}{$\sigma=0.5$} & &\multicolumn{3}{c}{$\sigma=1.5$}\\
 \cline{3-5}\cline{7-9}
\multicolumn{2}{c}{}&MSE&${\rm MSE_{diff}}$&PSE& &MSE&${\rm MSE_{diff}}$&PSE\\
\multicolumn{2}{c}{}&(sd)&(sd)&(sd)& &(sd)&(sd)&(sd)\\
\hline
%\\
\multirow{6}{*}{$\bm{\beta}_{1}^{*}$}&\multirow{2}{*}{BFL} &0.271&0.120&0.250&&1.528&0.715&1.171\\
&&(0.126)&(0.096)&(0.128)&&(0.718)&(0.537)&(0.519)\\
\cline{3-9}
%&& (2.202)&&&(0.360)&\\
%&BFNEG&&&\\
&\multirow{2}{*}{BFNEG}&0.183&0.094&0.196&&1.109&0.625&0.956\\
&&(0.094)&(0.101)&(0.116)&&(0.558)&(0.477)&(0.436)\\
\cline{3-9}
&\multirow{2}{*}{BFH}&\textbf{0.139}&\textbf{0.079}&\textbf{0.189}&&\textbf{0.622}&\textbf{0.540}&\textbf{0.682}\\
&&(0.084)&(0.085)&(0.121)&&(0.334)&(0.396)&(0.365)\\
%&& (1.055)&&&(0.127)&\\
\hline
\multirow{6}{*}{$\bm{\beta}_{2}^{*}$}&\multirow{2}{*}{BFL} &0.568& 0.209&0.627&&1.925&0.918&1.605\\
&&(0.275)&(0.156)&(0.382)&&(0.907)&(0.751)&(0.754)\\
\cline{3-9}
%&& (2.202)&&&(0.360)&\\
%&BFNEG&&&\\
&\multirow{2}{*}{BFNEG}&0.471&0.195&0.596&&1.297&0.703&1.266\\
&&(0.247)&(0.174)&(0.362)&&(0.832)&(0.544)&(0.729)\\
\cline{3-9}
&\multirow{2}{*}{BFH}&\textbf{0.413}&\textbf{0.176}&\textbf{0.593}&&\textbf{0.849}&\textbf{0.596}&\textbf{1.058}\\
&&(0.247)&(0.160)&(0.408)&&(0.524)&(0.649)&(0.631)\\
%&& (1.055)&&&(0.127)&\\
\hline
 \end{tabular}
\end{table}

\begin{table}[H]
  \centering
\caption{MSE (standard deviation), ${\rm MSE_{diff}}$, and PSE for Case 3. Bold font indicates smallest value among BFL, BFNEG, and BFH.}
\label{tab:case3}
\begin{tabular}{c|cccccccc}
\hline
\multicolumn{2}{c}{}&\multicolumn{3}{c}{$\sigma=0.5$} & &\multicolumn{3}{c}{$\sigma=1.5$}\\
\cline{3-5}\cline{7-9}
\multicolumn{2}{c}{}&MSE&${\rm MSE_{diff}}$&PSE& &MSE&${\rm MSE_{diff}}$&PSE\\
\multicolumn{2}{c}{}&(sd)&(sd)&(sd)& &(sd)&(sd)&(sd)\\
\hline
%\\
\multirow{6}{*}{$n=30$}&\multirow{2}{*}{BFL} &4.460&2.198&3.931&&14.295&5.098&8.940\\
&&(2.380)&(1.649)&(2.596)&&(5.784)&(3.278)&(3.904)\\
\cline{3-9}
%&& (2.202)&&&(0.360)&\\
%&BFNEG&&&\\
&\multirow{2}{*}{BFNEG}&\textbf{1.601}&\textbf{0.275}&\textbf{1.835}&&14.087&4.773&8.521\\
&&(0.300)&(0.148)&(0.486)&&(6.379)&(3.272)&(4.111)\\
\cline{3-9}
&\multirow{2}{*}{BFH}&1.623&0.736&2.408&&\textbf{7.478}&\textbf{3.119}&\textbf{5.530}\\
&&(0.848)&(0.556)&(2.656)&&(3.945)&(2.563)&(3.525)\\
%&& (1.055)&&&(0.127)&\\
\hline
\multirow{6}{*}{$n=50$}&\multirow{2}{*}{BFL} &1.805&0.533&1.648&&8.390&2.104&4.989\\
&&(0.642)&(0.410)&(1.094)&&(3.276)&(1.555)&(1.999)\\
\cline{3-9}
%&& (2.202)&&&(0.360)&\\
%&BFNEG&&&\\
&\multirow{2}{*}{BFNEG}&1.937&0.573&1.801&&10.584&2.514&6.250\\
&&(0.588)&(0.452)&(1.150)&&(4.257)&(1.970)&(2.583)\\
\cline{3-9}
&\multirow{2}{*}{BFH}&\textbf{0.787}&\textbf{0.298}&\textbf{1.424}&&\textbf{2.333}&\textbf{1.065}&\textbf{2.237}\\
&&(0.461)&(0.258)&(1.842)&&(1.273)&(0.832)&(1.909)\\
%&& (1.055)&&&(0.127)&\\
\hline
\end{tabular}
\end{table}
\begin{table}[H]
  \centering
\caption{MSE (standard deviation), ${\rm MSE_{diff}}$, and PSE for Case 4. Bold font indicates smallest value among BFL, BFNEG, and BFH.}
\label{tab:case4}
	\begin{tabular}{c|cccccccc}
  \hline
  \multicolumn{2}{c}{}&\multicolumn{3}{c}{$\sigma=0.5$} & &\multicolumn{3}{c}{$\sigma=1.5$}\\
  \cline{3-5}\cline{7-9}
 \multicolumn{2}{c}{}&MSE&${\rm MSE_{diff}}$&PSE& &MSE&${\rm MSE_{diff}}$&PSE\\
 \multicolumn{2}{c}{}&(sd)&(sd)&(sd)& &(sd)&(sd)&(sd)\\
 \hline
 %\\
 \multirow{6}{*}{$n=30$}&\multirow{2}{*}{BFL} &2.697&1.760&3.194&&8.537&4.341&9.160\\
 &&(1.221)&(1.293)&(1.488)&&(3.470)&(2.934)&(4.098)\\
 \cline{3-9}
 %&& (2.202)&&&(0.360)&\\
 %&BFNEG&&&\\
 &\multirow{2}{*}{BFNEG}&2.039&0.905&2.477&&9.716&4.830&9.897\\
 &&(0.809)&(0.702)&(1.045)&&(4.750)&(3.859)&(4.768)\\
 \cline{3-9}
 &\multirow{2}{*}{BFH}&\textbf{1.254}&\textbf{0.578}&\textbf{1.802}&&\textbf{4.290}&\textbf{2.452}&\textbf{4.998}\\
 &&(0.569)&(0.476)&(0.832)&&(2.906)&(2.848)&(2.516)\\
 %&& (1.055)&&&(0.127)&\\
 \hline
 \multirow{6}{*}{$n=50$}&\multirow{2}{*}{BFL} &1.468&0.503&1.457&&6.040&1.931&4.932\\
 &&(0.470)&(0.339)&(0.548)&&(2.511)&(1.376)&(1.940)\\
 \cline{3-9}
 %&& (2.202)&&&(0.360)&\\
 %&BFNEG&&&\\
 &\multirow{2}{*}{BFNEG}&1.708&0.554&1.722&&7.621&2.494&6.165\\
 &&(0.638)&(0.454)&(0.724)&&(3.335)&(1.891)&(2.519)\\
 \cline{3-9}
 &\multirow{2}{*}{BFH}&\textbf{0.716}&\textbf{0.260}&\textbf{1.067}&&\textbf{1.360}&\textbf{0.687}&\textbf{1.897}\\
 &&(0.355)&(0.211)&(0.561)&&(0.758)&(0.550)&(0.883)\\
 %&& (1.055)&&&(0.127)&\\
 \hline
\end{tabular}
\end{table}
The results are summarized in Tables 1, 2, 3, and 4. The proposed method BFH shows smaller MSEs and PSEs than BFL in almost all cases. This indicates that BFH outperformed BFL not only when $n>p$ but also when $n=p$ and $n<p$. In addition, BFH gives smaller ${\rm MSE_{diff}}$s than BFL in almost all cases. The reason is that BFH does not shrink non-zero differences of regression coefficients too much compared to BFL. Furthermore, BFH gives smaller values of MSE in 15 cases, PSE in 13 cases, and ${\rm MSE_{diff}}$ in 11 cases, out of the 16 cases, in comparison to BFNEG. BFH worked better than BFNEG especially in the case that successive regression coefficients had higher correlation than other pairs of regression coefficients. These results show that BFH gives a closer estimation to the true regression coefficients and can capture more non-zero differences of regression coefficients than the existing methods.\\

\subsection{Application}
We applied Bayesian HORSES with horseshoe prior in Section 3.2 to the Appalachian Mountains Soil Data dataset, which was analyzed in \citet{bondell2008simultaneous} and \citet{jang2015regression}. This dataset is available from \url{https://blogs.unimelb.edu.au/howard-bondell/#tab25} and was used for showing the relationship between soil characteristics and rich cove forest diversity. The dataset was collected at twenty 500 $\rm{m^{2}}$ plots in the Appalachian Mountains. Forest diversity, which is represented as the number of different plant species, is used for response variables and 15 soil characteristics in 20 plots are used as explanatory variables. The data are the average of five equally spaced measurements in each plot. We standardized soil data before the analysis.\\
\indent We compared Bayesian HORSES with horseshoe prior (BHH) to Bayesian HORSES (BH) and Bayesian HORSES with NEG prior (BHNEG). For this application, we chose the hyper-parameters $\lambda_{2}^{2}$ for BH from five candidates between $10^{-4}$ and $10^{-2}$, $\tilde{\tau}^{2}$ for BHH from five candidates between $10^{4}$ and $10^{6}$, and $\gamma_{2}^{2}$ for BHNEG from five candidates between 1 and 10. We set the hyper-parameter $\lambda_{2}$ for BHNEG as 0.5. We executed a leave-one-out cross-validation to assess the performance of the models. In each estimation, the Gibbs sampling was run with 10,000 iterations and 5,000 iterations were discarded as burn-in. The mean values of cross-validation, CV, are summarized in Table \ref{tab:case5}.
From Table \ref{tab:case5}, the value of CV for our proposed method is smaller than that for BH. BHNEG gives the smallest value of CV, but gives the largest value of standard deviation. On the other hand, our proposed method gives the smallest value of standard deviation and the second largest value of CV. Table \ref{tab:case6} shows the regression coefficients estimated by all 20 samples. Note that the hyper-parameters were selected by WAIC. The scales of the estimated values are different from those of the results of \citet{jang2015regression}, but we observe that BHH captures almost the same group structure as found in \citet{jang2015regression}. Considering these results, BHH gives relatively stable estimation capturing the group structure of variables compared to the existing methods.
\begin{table}[H]
  \centering
\caption{Results for analysis of Appalachian Mountains Soil Data.}
\label{tab:case5}
	\begin{tabular}{cccc}
  \hline
  &BH&BHNEG&BHH\\
	\hline
	CV& 0.000678&0.000619&0.000656\\
	(sd)&(0.000840)&(0.000852)&(0.000805)\\
 %&& (1.055)&&&(0.127)&\\
 \hline
\end{tabular}
\end{table}

\begin{table}[H]
  \centering
\caption{Estimated values of regression coefficients in analysis of Appalachian Mountains Soil Data.}
\label{tab:case6}
	\begin{tabular}{lS[group-digits=false,table-format=+1.5]S[group-digits=false,table-format=+1.5]S[group-digits=false,table-format=+1.5]S[table-format=+1.4]}
  \hline
  &{BH}&{BHNEG}&{BHH}\\
	\hline%-0.00441
	Base saturation&-0.00441 &-0.00199& -0.00330\\
	Sum cations&-0.00745&-0.00419&  -0.00638\\
	CEC buffer&-0.00501&-0.00570&-0.00599\\
	Calcium&-0.00916 &-0.00505& -0.00702\\
	Magnesium&0.00306 &-0.00120& -0.00003\\
	Potassium&-0.00799 &-0.00378& -0.00540\\
	Sodium&0.00003  &0.00020& 0.00035\\
	Phosphorus&0.00781 &0.00355& 0.00609\\
	Copper&0.01489&0.00646& 0.01132\\
	Zinc&-0.00586&0.00107& -0.00194\\
	Manganese&0.01077 &0.00789& 0.00946\\
	Humic matter&-0.02309 &-0.01444& -0.01930\\
	Density&-0.00089 &0.00232& 0.00060 \\
	pH&0.01156 &0.00388& 0.00682\\
  Exchangeable acidity&0.00580&-0.00132&0.00105\\
 \hline
\end{tabular}
\end{table}
\newpage
\section{Conclusions}
\label{sec:conclusion}
We proposed Bayesian fused lasso modeling with horseshoe prior, and then developed the Gibbs sampler for the parameters by using a scale mixture of normals and a hierarchical expression of a half-Cauchy prior. In addition, we extended the method to the Bayesian HORSES. Through numerical studies, we showed our proposed method is superior to the existing methods in terms of prediction and estimation accuracy. \\
\indent In Bayesian HORSES with horseshoe prior, we select the value of global shrinkage parameter $\tilde{\tau}^2$ by WAIC. It would be interesting to assume any proper prior on $\tilde{\tau}^2$. Considering how to accelerate the convergence of the Gibbs sampling for our proposed method would also be interesting. We leave these topics as future work.

\section*{Acknowledgements}
	S. K. was supported by JSPS KAKENHI Grant Numbers JP19K11854 and JP20H02227.

\appendix

\section{Detailed calculation of the formula \eqref{eq:yukoprior_rewrite}}
The detailed calculation of rewriting \eqref{eq:yukoprior} as \eqref{eq:yukoprior_rewrite} is as follows:
\begin{equation}
	\begin{split}
		\pi(\bm{\beta}\mid\sigma^{2})
		 \propto{}&(\sigma^{2})^{-\frac{p}{2}}\prod_{j=1}^{p}\int \frac{1}{\sqrt{2\pi\tau_{j}^{2}}}\exp\left(-\frac{\beta_{j}^{2}}{2\sigma^{2}\tau_{j}^{2}}\right)\frac{\tilde{\lambda}_{1}^{2}}{2}\exp\left(-\frac{\tilde{\lambda}_{1}^{2}}{2}\tau_{j}^{2}\right)d\tau_{j}^{2}
	  \\&\times\prod_{j=2}^{p} \iint\left[\frac{1}{\sqrt{2\pi\lambda_{j}^{2}\tilde{\tau}^{2}\sigma^{2}}}\exp\left\{-\frac{(\beta_{j}-\beta_{j-1})^2}{2\lambda_{j}^{2}\tilde{\tau}^{2}\sigma^{2}}\right\}\right.
	  \\&
	  \times\int\frac{(\frac{1}{\nu_{j}})^{\frac{1}{2}}}{\Gamma(\frac{1}{2})}(\lambda_{j}^{2})^{-\frac{3}{2}}\exp\left(-\frac{1}{\nu_{j}\lambda_{j}^{2}}\right)\frac{1}{\Gamma(\frac{1}{2})}(\nu_{j})^{-\frac{3}{2}}\exp\left(-\frac{1}{\nu_{j}}\right)d\nu_j
	  \\&
	  \times\int\left.\frac{(\frac{1}{\xi})^{\frac{1}{2}}}{\Gamma(\frac{1}{2})}(\tilde{\tau}^{2})^{-\frac{3}{2}}\exp\left(-\frac{1}{\xi\tilde{\tau}^{2}}\right)\frac{1}{\Gamma(\frac{1}{2})}(\xi)^{-\frac{3}{2}}\exp\left(-\frac{1}{\xi}\right)d\xi\right] d\lambda_{j}^{2}d\tilde{\tau}^{2}
	  \\\propto{}&%2行目
	 \int \ldots \int \prod_{j=1}^{p}\frac{1}{\sqrt{2\pi\sigma^{2}\tau_{j}^{2}}}\exp\left(-\frac{\beta_{j}^{2}}{2\sigma^{2}\tau_{j}^{2}}\right) \prod_{j=1}^{p}\frac{\tilde{\lambda}_{1}^{2}}{2}\exp\left(-\frac{\tilde{\lambda}_{1}^{2}}{2}\tau_{j}^{2}\right)
	 \\&\times\prod_{j=2}^{p}\frac{1}{\sqrt{2\pi\lambda_{j}^{2}\tilde{\tau}^{2}\sigma^{2}}}\exp\left\{-\frac{(\beta_{j}-\beta_{j-1})^2}{2\lambda_{j}^{2}\tilde{\tau}^{2}\sigma^{2}}\right\}\prod_{j=2}^{p}\frac{(\frac{1}{\nu_{j}})^{\frac{1}{2}}}{\Gamma(\frac{1}{2})}(\lambda_{j}^{2})^{-\frac{3}{2}}\exp\left(-\frac{1}{\nu_{j}\lambda_{j}^{2}}\right)
	  \\& \times\frac{(\frac{1}{\xi})^{\frac{1}{2}}}{\Gamma(\frac{1}{2})}(\tilde{\tau}^{2})^{-\frac{3}{2}}\exp(-\frac{1}{\xi\tilde{\tau}^{2}})\frac{1}{\Gamma(\frac{1}{2})}(\xi)^{-\frac{3}{2}}\exp\left(-\frac{1}{\xi}\right)\prod_{j=2}^{p}\frac{1}{\Gamma(\frac{1}{2})}(\nu_{j})^{-\frac{3}{2}}\exp\left(-\frac{1}{\nu_{j}}\right)
	  \\&\times d\tilde{\tau}^{2}d\xi\prod_{j=1}^{p}d\tau_{j}^{2}\prod_{j=2}^{p}d\lambda_{j}^{2}\prod_{j=2}^{p}d\nu_{j}\\
		\propto{}&%3行目
		\int \ldots \int (\sigma^{2})^{-\frac{2p-1}{2}}(\tilde{\tau}^{2})^{-\frac{p-1}{2}}\pi(\tilde{\tau}^{2}\mid\xi)\pi(\xi)\prod_{j=1}^{p}(\tau_{j}^{2})^{-\frac{1}{2}}\prod_{j=2}^{p}(\lambda_{j}^{2})^{-\frac{1}{2}}\exp\left(-\frac{1}{2\sigma^{2}}\bm{\beta}^{T}\bm{B}^{-1}\bm{\beta}\right)
		 \\& \times\prod_{j=1}^{p}\pi(\tau_{j}^{2})\prod_{j=2}^{p}\pi(\lambda_{j}^{2}\mid\nu_{j})\prod_{j=2}^{p}\pi(\nu_{j})d\tilde{\tau}^{2}d\xi\prod_{j=1}^{p}d\tau_{j}^{2}\prod_{j=2}^{p}d\lambda_{j}^{2}\prod_{j=2}^{p}d\nu_{j}.
		\end{split}\nonumber
\end{equation}
\bibliographystyle{kawano_apa}
\bibliography{soturonsanko}

\end{document}